\newenvironment{Cscope%
}{}{}%
\newenvironment{centre}{\begin{center}}{\end{center}}\documentstyle[12pt]%
{article}\newlength\Cscr\newlength
\Csave\newlength\Ctenthex\newlength
\CFxsize\newlength\CFxsizeps\newlength\CFsizemakebox\newlength
\CFleftcrop\newlength\CFrightcrop\newlength\CZtbldist\newlength
\CZfigdist
\newlength\CGDnum\newlength\CGDtext\newcounter{Cscr}\newcounter{%
Csave}\newcounter{CBcit}%
\newcounter{CBauthornu%
m}\newcounter{Ceqindent}\newcounter{CBtnc}
\newcounter{CBtntc}\newcounter{CEht}%
\newcounter{CbsA}\newcounter{CbsB}\newcounter
{CbsC}\newcounter{CbsD}\hoffset\Cscr\voffset
\Cscr\protect
\begin{document}\renewcommand\theequation{\arabic{e%
quation}}\renewcommand\thetable{\arabic{table}}\renewcommand
\thefigure{\arabic{figure}}\renewcommand\thesection{\Roman{secti%
on}}\renewcommand\thesubsection{\Alph{subsection}}\renewcommand
\thesubsubsection{\arabic{subsubsection}}\setcounter{CEht}{10}%
\setcounter{CbsA}{1}\setcounter{CbsB}{1}\setcounter{CbsC}{1}%
\setcounter{CbsD}{1}\hfill UM--P--97/19\par\hfill To appear in t%
he \mbox{}\protect\/{\protect\em American Journal of Physics%
\protect\/}\par{\centering\protect\mbox{}\\*[\baselineskip]{%
\large\bf Classical antiparticles}\\*}\addtocounter{CBtntc}{1}%
\addtocounter{CBtntc}{1}\addtocounter{CBtntc}{1}{\centering
\protect\mbox{}\\John P.~Costella,$^{\fnsymbol{CBtnc}}$%
\addtocounter{CBtnc}{1}\ Bruce H.~J.~McKellar,$^{\fnsymbol{CBtnc%
}}$\addtocounter{CBtnc}{1}\ and Andrew A.~Rawlinson$^{\fnsymbol{%
CBtnc}}$\addtocounter{CBtnc}{1}\\*}{\centering{\small\mbox{}%
\protect\/{\protect\em School of Physics, The University of Melb%
ourne, Parkville, Victoria 3052, Australia\protect\/}}\\}{%
\centering\protect\mbox{}\\(29 March 1997)\\} \par\vspace
\baselineskip\begin{centre}{\small\bf Abstract}\end{centre}%
\vspace{-1.25ex}\vspace{-0.75\baselineskip}\par\setlength{\Csave
}{\parskip}\begin{quote}\setlength{\parskip}{0.5\baselineskip}%
\small\noindent We review how \mbox{}\protect\/{\protect\em anti%
particles\protect\/} may be introduced in classical relativistic 
mechanics, and emphasize that many of their paradoxical properti%
es can be more transparently understood in the classical than in 
the quantum domain. \end{quote}\setlength{\parskip}{\Csave}\par
\refstepcounter{section}\vspace{1.5\baselineskip}\par{\centering
\bf\thesection. Introduction\\*[0.5\baselineskip]}\protect\indent
\label{sect:Intro}Recently,$^{\ref{cit:Costella1995}}$ we review%
ed briefly the physics and early history of the Foldy--Wout\-huy%
sen\ transformation,$^{}$$^{\ref{cit:Newton1949}}$$^{,}$$^{\ref{%
cit:Foldy1950}}${} emphasizing that the transformed representati%
on is the only one in which a \mbox{}\protect\/{\protect\em clas%
sical limit\protect\/} of the Dirac equation can be meaningfully 
extracted, in terms of particles and antiparticles. But few text%
books actually describe how antiparticles \mbox{}\protect\/{%
\protect\em can\protect\/} be dealt with in classical mechanics. 
Discussions of antiparticles usually begin with the ``negative e%
nergy problem'': the inevitable introduction, in relativistic me%
chanics, of what appears to be a ``spurious'' set of mirror eige%
nstates of negative energy; their re\-interpretation by Dirac as 
``holes'' in a filled Fermi sea of vacuum electrons; and their f%
urther reformulation, in quantum field theory, as completely val%
id eigenstates in their own right. But this introduction is alto%
gether too late: while its appearance in a course on relativisti%
c quantum mechanics reflects accurately the \mbox{}\protect\/{%
\protect\em historical\protect\/} development of the theory of a%
ntiparticles, it can tend to hide completely the fact that it is 
\mbox{}\protect\/{\protect\em relativistic mechanics itself%
\protect\/} that makes possible the phenomenon of antiparticle m%
otion---quantum mechanics is by no means a prerequisite.\par Arg%
uably, a thorough preliminary understanding of the \mbox{}%
\protect\/{\protect\em classical\protect\/} theory of antipartic%
les better equips the student for tackling the same issues when 
they arise in relation to the Dirac equation. It is this topic t%
hat we shall review in this paper.\par\refstepcounter{section}%
\vspace{1.5\baselineskip}\par{\centering\bf\thesection. The prop%
er time\\*[0.5\baselineskip]}\protect\indent\label{sect:Tau}Cons%
ider a structureless point particle. Classically, its kinematica%
l state at any time $t$ consists simply of the three components 
of its \mbox{}\protect\/{\protect\em three-position\protect\/} $%
{\protect\mbox{\protect\boldmath{$z$}}}(t)$. We assume that ${%
\protect\mbox{\protect\boldmath{$z$}}}(t)$ is a continuous funct%
ion of $t$ that is sufficiently differentiable for our purposes. 
In special relativity, we form the \mbox{}\protect\/{\protect\em
four-position\protect\/} $z^\mu$ of the particle: \setcounter{Ce%
qindent}{0}\protect\begin{eqnarray}\hspace{-1.3ex}&\displaystyle
z^\mu\equiv(t,{\protect\mbox{\protect\boldmath{$z$}}}),\protect
\nonumber\setlength{\Cscr}{\value{CEht}\Ctenthex}\addtolength{%
\Cscr}{-1.0ex}\protect\raisebox{0ex}[\value{CEht}\Ctenthex][\Cscr
]{}\protect\end{eqnarray}\setcounter{CEht}{10}where we shall alw%
ays use units in which \mbox{$c=1$}. The continuous function ${%
\protect\mbox{\protect\boldmath{$z$}}}(t)$ specifies the \mbox{}%
\protect\/{\protect\em path\protect\/} of the particle in Minkow%
ski spacetime---its \mbox{}\protect\/{\protect\em worldline%
\protect\/}. To parame\-ter\-ize its ``length'', in a Lorentz-in%
variant way, we consider an infinitesimal differential element o%
f the path, \setcounter{Ceqindent}{0}\protect\begin{eqnarray}%
\hspace{-1.3ex}&\displaystyle dz^\mu(t)\equiv(dt,d{\protect\mbox
{\protect\boldmath{$z$}}}(t)),\protect\nonumber\setlength{\Cscr}%
{\value{CEht}\Ctenthex}\addtolength{\Cscr}{-1.0ex}\protect
\raisebox{0ex}[\value{CEht}\Ctenthex][\Cscr]{}\protect\end{eqnar%
ray}\setcounter{CEht}{10}where $d{\protect\mbox{\protect\boldmath
{$z$}}}(t)$ is the infinitesimal change in position ${\protect
\mbox{\protect\boldmath{$z$}}}(t)$ in the infinitesimal time int%
erval from $t$ to \mbox{$t+dt$}. We now consider the Lorentz-inv%
ariant quantity \setcounter{Ceqindent}{0}\protect\begin{eqnarray%
}\protect\left.\protect\begin{array}{rcl}\protect\displaystyle
\hspace{-1.3ex}&\protect\displaystyle d\tau^2(t)\equiv dz^\mu(t)%
dz_\mu(t)\equiv dt^2-d{\protect\mbox{\protect\boldmath{$z$}}}^2(%
t),\setlength{\Cscr}{\value{CEht}\Ctenthex}\addtolength{\Cscr}{-%
1.0ex}\protect\raisebox{0ex}[\value{CEht}\Ctenthex][\Cscr]{}%
\protect\end{array}\protect\right.\protect\label{eq:Tau-DTauSqua%
red}\protect\end{eqnarray}\setcounter{CEht}{10}where we employ a 
\mbox{$(\mbox{$+$},\mbox{$-$},\mbox{$-$},\mbox{$-$})$} metric. W%
hat we would \mbox{}\protect\/{\protect\em like\protect\/} to do 
is define a quantity $d\tau(t)$ that would provide a measure of 
``length'' along the worldline. But the Lorentz-invariant expres%
sion (\protect\ref{eq:Tau-DTauSquared}) involves not $d\tau$, bu%
t rather the \mbox{}\protect\/{\protect\em square\protect\/} of 
$d\tau$. Thus, $d\tau$ can only be defined \mbox{}\protect\/{%
\protect\em up to a sign\protect\/}: \setcounter{Ceqindent}{0}%
\protect\begin{eqnarray}\protect\left.\protect\begin{array}{rcl}%
\protect\displaystyle\hspace{-1.3ex}&\protect\displaystyle d\tau
\equiv\pm\sqrt{dz^\mu dz_\mu}.\setlength{\Cscr}{\value{CEht}%
\Ctenthex}\addtolength{\Cscr}{-1.0ex}\protect\raisebox{0ex}[%
\value{CEht}\Ctenthex][\Cscr]{}\protect\end{array}\protect\right
.\protect\label{eq:Tau-DTau}\protect\end{eqnarray}\setcounter{CE%
ht}{10}To investigate the meaning of this ambiguity in the sense 
of $d\tau$, let us consider the special case in which the partic%
le is instantaneously at rest, with respect to our own inertial 
co\-ordinate system: \setcounter{Ceqindent}{0}\protect\begin{eqn%
array}\hspace{-1.3ex}&\displaystyle d{\protect\mbox{\protect
\boldmath{$z$}}}=\mbox{\bf0}.\protect\nonumber\setlength{\Cscr}{%
\value{CEht}\Ctenthex}\addtolength{\Cscr}{-1.0ex}\protect
\raisebox{0ex}[\value{CEht}\Ctenthex][\Cscr]{}\protect\end{eqnar%
ray}\setcounter{CEht}{10}In this case, we find \setcounter{Ceqin%
dent}{0}\protect\begin{eqnarray}\hspace{-1.3ex}&\displaystyle d%
\tau=\pm dt.\protect\nonumber\setlength{\Cscr}{\value{CEht}%
\Ctenthex}\addtolength{\Cscr}{-1.0ex}\protect\raisebox{0ex}[%
\value{CEht}\Ctenthex][\Cscr]{}\protect\end{eqnarray}\setcounter
{CEht}{10}The solution \mbox{$d\tau=dt$} for a particle at rest 
is the one usually presented in introductory texts on special re%
lativity: such a $d\tau$ is obviously equal to the passage of ti%
me as measured in the \mbox{}\protect\/{\protect\em instantaneou%
s rest frame\protect\/} of the particle. For a particle undergoi%
ng arbitrary relativistic motion, we assume that the particle it%
self possesses its own ``cumulative time'' or ``age'', which we 
term the \mbox{}\protect\/{\protect\em proper time\protect\/}, t%
hat can be calculated by summing up all of the $d\tau$ along its 
worldline: \setcounter{Ceqindent}{0}\protect\begin{eqnarray}%
\hspace{-1.3ex}&\displaystyle\tau({\cal E})\equiv\hspace{-0.5mm}%
\protect\mbox{}\hspace{-0.1mm}\protect\int_{{\cal E}_0}^{{\cal E%
}}\protect\mbox{}\hspace{-0.5mm}\hspace{-0.6mm}d\tau,\protect
\nonumber\setlength{\Cscr}{\value{CEht}\Ctenthex}\addtolength{%
\Cscr}{-1.0ex}\protect\raisebox{0ex}[\value{CEht}\Ctenthex][\Cscr
]{}\protect\end{eqnarray}\setcounter{CEht}{10}where $\tau({\cal
E})$ is the proper time at event ${\cal E}$ on the worldline, an%
d where the event ${\cal E}_0$ on the worldline defines the (arb%
itrary) origin of $\tau$. Since the worldline of any classical p%
article passes through each constant-$t$ hyperplane once and onl%
y once, we can replace the events ${\cal E}$ and ${\cal E}_0$ by 
their corresponding co\-ordinate times $t$ and $t_0$, and hence 
determine $\tau$ as a function of $t$: \setcounter{Ceqindent}{0}%
\protect\begin{eqnarray}\protect\left.\protect\begin{array}{rcl}%
\protect\displaystyle\hspace{-1.3ex}&\protect\displaystyle\tau(t%
)=\hspace{-0.5mm}\protect\mbox{}\hspace{-0.1mm}\protect\int_{t_0%
}^{t}\protect\mbox{}\hspace{-0.5mm}\hspace{-0.6mm}dt'\,\mbox{$%
\protect\displaystyle\protect\frac{d\tau}{dt'}$}=\hspace{-0.5mm}%
\protect\mbox{}\hspace{-0.1mm}\protect\int_{t_0}^{t}\protect\mbox
{}\hspace{-0.5mm}\hspace{-0.6mm}dt'\,\mbox{$\protect\displaystyle
\protect\frac{1}{\gamma(t')}$},\setlength{\Cscr}{\value{CEht}%
\Ctenthex}\addtolength{\Cscr}{-1.0ex}\protect\raisebox{0ex}[%
\value{CEht}\Ctenthex][\Cscr]{}\protect\end{array}\protect\right
.\protect\label{eq:Tau-TauOfT}\protect\end{eqnarray}\setcounter{%
CEht}{10}where we have made use of Eq.~(\protect\ref{eq:Tau-DTau%
Squared}): \setcounter{Ceqindent}{0}\protect\begin{eqnarray}%
\hspace{-1.3ex}&\displaystyle\mbox{$\protect\displaystyle\protect
\frac{d\tau(t)}{dt}$}=\sqrt{1-\protect\left(\mbox{$\protect
\displaystyle\protect\frac{d{\protect\mbox{\protect\boldmath{$z$%
}}}(t)}{dt}$}\protect\right)^{\!\!2}}\equiv\sqrt{1-{\protect\mbox
{\protect\boldmath{$v$}}}^2(t)}\equiv\mbox{$\protect\displaystyle
\protect\frac{1}{\gamma(t)}$}.\protect\nonumber\setlength{\Cscr}%
{\value{CEht}\Ctenthex}\addtolength{\Cscr}{-1.0ex}\protect
\raisebox{0ex}[\value{CEht}\Ctenthex][\Cscr]{}\protect\end{eqnar%
ray}\setcounter{CEht}{10}The standard textbook result (\protect
\ref{eq:Tau-TauOfT}) shows that when the speed $v$ of the partic%
le is much smaller than the speed of light, the factor $\gamma(t%
)$ is close to unity, and the passage of proper time is indistin%
guishable from that of co\-ordinate time; but if the particle's 
motion is such that its speed rises to an appreciable fraction o%
f the speed of light, the factor $\gamma(t)$ rises above unity, 
and the particle ``ages'' more slowly. In all cases, however, th%
e particle \mbox{}\protect\/{\protect\em gets older\protect\/}: 
special relativity only seems to modify the rate; it warps our v%
iew of the world, but it does not throw it into reverse.\par
\refstepcounter{section}\vspace{1.5\baselineskip}\par{\centering
\bf\thesection. Classical antiparticles\\*[0.5\baselineskip]}%
\protect\indent\label{sect:Antiparticles}Let us now consider the 
\mbox{}\protect\/{\protect\em other\protect\/} solution in Eq.~(%
\protect\ref{eq:Tau-DTau}) for a particle at rest, namely, 
\setcounter{Ceqindent}{0}\protect\begin{eqnarray}\protect\left.%
\protect\begin{array}{rcl}\protect\displaystyle\hspace{-1.3ex}&%
\protect\displaystyle d\tau=&\hspace{-1.3ex}\protect\displaystyle
-dt.\setlength{\Cscr}{\value{CEht}\Ctenthex}\addtolength{\Cscr}{%
-1.0ex}\protect\raisebox{0ex}[\value{CEht}\Ctenthex][\Cscr]{}%
\protect\end{array}\protect\right.\protect\label{eq:Antiparticle%
s-TimeBackwards}\protect\end{eqnarray}\setcounter{CEht}{10}Even 
to a student possessing a good knowledge of special relativity, 
Eq.~(\protect\ref{eq:Antiparticles-TimeBackwards}) does not look 
familiar at all. It seems to imply that a particle at rest with 
respect to our Lorentz co\-ordinate system might somehow believe 
that time evolves in the \mbox{}\protect\/{\protect\em opposite 
direction\protect\/} to what we do! For example, if we determine 
that some spacetime event ${\cal E}'$ is definitely \mbox{}%
\protect\/{\protect\em earlier\protect\/} than another event ${%
\cal E}''$ (i.e., ${\cal E}'$ lies within the backward lightcone 
of ${\cal E}''$), then a particle whose own ``proper time'' obey%
s (\protect\ref{eq:Antiparticles-TimeBackwards}) would insist, t%
o the contrary, that ${\cal E}'$ is definitely \mbox{}\protect\/%
{\protect\em later\protect\/} than ${\cal E}''$ (i.e., from the 
particle's point of view, ${\cal E}'$ lies within the \mbox{}%
\protect\/{\protect\em forward\protect\/} lightcone of ${\cal E}%
''$).\par The problem is that, by the principles of relativity, 
\mbox{}\protect\/{\protect\em such a particle is just as valid a%
n observer of the universe as we are\protect\/}: it agrees with 
us that the speed of light is unity in all inertial frames. We h%
ave no physically acceptable justification for dismissing its co%
unterintuitive view of the world. We must conclude that \mbox{}%
\protect\/{\protect\em both\protect\/} of the solutions (\protect
\ref{eq:Tau-DTau}) are equally valid definitions of the passage 
of proper time. (This is analogous to the fact$^{\ref{cit:Jackso%
n1975}}$ that both the retarded \mbox{}\protect\/{\protect\em an%
d\protect\/} the advanced Li\'enard--Wiechert potentials for a p%
oint charge are equally valid solutions of Maxwell's equations.)%
\par Armed with a thorough knowledge of relativistic quantum mec%
hanics and quantum field theory, Stueckelberg$^{\ref{cit:Stuecke%
lberg1942}}$ and Feynman$^{}$$^{\ref{cit:Feynman1948}}$$^{,}$$^{%
\ref{cit:Feynman1949}}$$^{,}$$^{\ref{cit:Feynman1987}}${} made t%
he following realization: a particle for which $d\tau$ evolves i%
n the opposite sense to the $dt$ in our particular Lorentz frame 
of reference is simply in \mbox{}\protect\/{\protect\em antipart%
icle motion\protect\/} with respect to us. Of course, there are 
no classical forces that can change ``particle motion'' into ``a%
ntiparticle motion''---the two regimes are as disjoint as the in%
teriors of the forward and backward lightcones; but, even classi%
cally, this does not bar the possibility that a particle might h%
ave \mbox{}\protect\/{\protect\em always\protect\/} been in anti%
particle motion.\par Following this argument to its logical conc%
lusion, it could be noted that there is a similar ambiguity of s%
ign when parame\-ter\-izing path lengths in \mbox{}\protect\/{%
\protect\em Euclidean\protect\/} space, since there the invarian%
t interval is also squared: \setcounter{Ceqindent}{0}\protect
\begin{eqnarray}\hspace{-1.3ex}&\displaystyle dl^2\equiv d{%
\protect\mbox{\protect\boldmath{$z$}}}^2,\protect\nonumber
\setlength{\Cscr}{\value{CEht}\Ctenthex}\addtolength{\Cscr}{-1.0%
ex}\protect\raisebox{0ex}[\value{CEht}\Ctenthex][\Cscr]{}\protect
\end{eqnarray}\setcounter{CEht}{10}and hence we could equally we%
ll measure length one way along the path, or in the opposite way%
. But we are \mbox{}\protect\/{\protect\em already\protect\/} us%
ed to the idea that, at a fundamental level, travel\-ing to the 
left is no more difficult than travel\-ing to the right. The cru%
cial difference in Minkowski space is precisely the fact that cl%
assical forces \mbox{}\protect\/{\protect\em do not\protect\/} r%
everse the sense in which ``time is traversed''; our intuition w%
ith Galilean mechanics is rooted firmly in the belief that every%
one agrees on the direction that time is travel\-ing. Relativist%
ically boosting to another frame of reference ``warps'' the rate 
at which clocks tick, but it does not reverse it; in contrast, t%
ime-reversal is a \mbox{}\protect\/{\protect\em discrete\protect
\/} symmetry, and cannot be brought into contact with ``intuitiv%
e'' physics by a continuous transformation.\par\refstepcounter{s%
ection}\vspace{1.5\baselineskip}\par{\centering\bf\thesection. $%
{\cal C}$, ${\cal P}$, and ${\cal T}$\\*[0.5\baselineskip]}%
\protect\indent\label{sect:CPT}Another way of recognizing the po%
ssibility of the existence of antiparticle motion, in any relati%
vistically complete theory of mechanics, is to consider the fund%
amental symmetries of the \mbox{}\protect\/{\protect\em Lorentz 
group\protect\/}---namely, those transformations under which the 
interval (\protect\ref{eq:Tau-DTauSquared}) is invariant. Intuit%
ive, introductory constructions of the proper time generally mak%
e use of everyday objects, such as people, trains, measuring rod%
s, clocks, and so on. In thinking about such everyday objects---%
even in relativistic terms---we usually only consider \mbox{}%
\protect\/{\protect\em proper\protect\/} Lorentz transformations 
(boosts and rotations)---or, at most, \mbox{}\protect\/{\protect
\em orthochronous\protect\/} ones (proper transformations with o%
r without the parity transformation). But the interval (\protect
\ref{eq:Tau-DTauSquared}) is also invariant under \mbox{}\protect
\/{\protect\em non-orthochronous\protect\/} Lorentz transformati%
ons---those involving the time-reversal operation---and it is pr%
ecisely such transformations that convert what appears to be ``n%
ormal'' particle motion into ``antiparticle'' motion.\par Let us 
make this argument more concrete. Consider the \mbox{}\protect\/%
{\protect\em parity\protect\/} operation, which in classical phy%
sics simply changes the sign of the spatial co\-ordinates in a g%
iven Lorentz frame: \setcounter{Ceqindent}{0}\protect\begin{eqna%
rray}\hspace{-1.3ex}&\displaystyle{\cal P}:\;{\protect\mbox{%
\protect\boldmath{$x$}}}\rightarrow-{\protect\mbox{\protect
\boldmath{$x$}}}.\protect\nonumber\setlength{\Cscr}{\value{CEht}%
\Ctenthex}\addtolength{\Cscr}{-1.0ex}\protect\raisebox{0ex}[%
\value{CEht}\Ctenthex][\Cscr]{}\protect\end{eqnarray}\setcounter
{CEht}{10}The \mbox{}\protect\/{\protect\em time-reversal\protect
\/} operation does likewise for the time co\-ordinate: 
\setcounter{Ceqindent}{0}\protect\begin{eqnarray}\hspace{-1.3ex}%
&\displaystyle{\cal T}:\;t\rightarrow-t.\protect\nonumber
\setlength{\Cscr}{\value{CEht}\Ctenthex}\addtolength{\Cscr}{-1.0%
ex}\protect\raisebox{0ex}[\value{CEht}\Ctenthex][\Cscr]{}\protect
\end{eqnarray}\setcounter{CEht}{10}Under the combined operations 
of ${\cal P}$ and ${\cal T}$, all four components of $dz^\mu$ ar%
e reversed in sign: \setcounter{Ceqindent}{0}\protect\begin{eqna%
rray}\protect\left.\protect\begin{array}{rcl}\protect
\displaystyle\hspace{-1.3ex}&\protect\displaystyle{\cal P}{\cal
T}:\;dz^\mu\rightarrow-dz^\mu.\setlength{\Cscr}{\value{CEht}%
\Ctenthex}\addtolength{\Cscr}{-1.0ex}\protect\raisebox{0ex}[%
\value{CEht}\Ctenthex][\Cscr]{}\protect\end{array}\protect\right
.\protect\label{eq:CPT-PTz}\protect\end{eqnarray}\setcounter{CEh%
t}{10}The three-velocity, being a ratio of the spatial part $d{%
\protect\mbox{\protect\boldmath{$z$}}}$ to the temporal part $dz%
^0$, is therefore unchanged: \setcounter{Ceqindent}{0}\protect
\begin{eqnarray}\hspace{-1.3ex}&\displaystyle{\cal P}{\cal T}:\;%
{\protect\mbox{\protect\boldmath{$v$}}}\equiv\mbox{$\protect
\displaystyle\protect\frac{d{\protect\mbox{\protect\boldmath{$z$%
}}}}{dt}$}\rightarrow{\protect\mbox{\protect\boldmath{$v$}}}.%
\protect\nonumber\setlength{\Cscr}{\value{CEht}\Ctenthex}%
\addtolength{\Cscr}{-1.0ex}\protect\raisebox{0ex}[\value{CEht}%
\Ctenthex][\Cscr]{}\protect\end{eqnarray}\setcounter{CEht}{10}Le%
t us now \mbox{}\protect\/{\protect\em try\protect\/} to define 
proper time so that the sign of $d\tau$ is always taken in the s%
ame sense as $dt$. Consider a free particle, that has a three-ve%
locity ${\protect\mbox{\protect\boldmath{$v$}}}$ in a given Lore%
ntz frame of reference; and let us define \setcounter{Ceqindent}%
{0}\protect\begin{eqnarray}\protect\left.\protect\begin{array}{r%
cl}\protect\displaystyle\hspace{-1.3ex}&\protect\displaystyle d%
\tau=+\mbox{$\protect\displaystyle\protect\frac{dt}{\gamma}$},%
\setlength{\Cscr}{\value{CEht}\Ctenthex}\addtolength{\Cscr}{-1.0%
ex}\protect\raisebox{0ex}[\value{CEht}\Ctenthex][\Cscr]{}\protect
\end{array}\protect\right.\protect\label{eq:CPT-TauPos}\protect
\end{eqnarray}\setcounter{CEht}{10}where we always define $\gamma
$ as the \mbox{}\protect\/{\protect\em positive\protect\/} squar%
e-root: \setcounter{Ceqindent}{0}\protect\begin{eqnarray}\hspace
{-1.3ex}&\displaystyle\gamma\equiv\gamma(v)\equiv\mbox{$\protect
\displaystyle\protect\frac{1}{\sqrt{1-v^2}}$}.\protect\nonumber
\setlength{\Cscr}{\value{CEht}\Ctenthex}\addtolength{\Cscr}{-1.0%
ex}\protect\raisebox{0ex}[\value{CEht}\Ctenthex][\Cscr]{}\protect
\end{eqnarray}\setcounter{CEht}{10}The choice of sign (\protect
\ref{eq:CPT-TauPos}) lets us label the worldline with values of 
$\tau$ in the ``standard'' way, such that the $\tau$ values incr%
ease in the direction of increasing co\-ordinate time $t$.\par I%
f we now apply the operation ${\cal P}{\cal T}$ to the above Lor%
entz frame, we are placed in a new, equally valid Lorentz frame, 
in which all directions---space \mbox{}\protect\/{\protect\em an%
d\protect\/} time---have been reversed. But this operation does 
not affect our $\tau$ markings on the particle's worldline, sinc%
e the proper time is a property of the particle itself. Thus, wi%
th respect to the \mbox{}\protect\/{\protect\em new\protect\/} L%
orentz frame, $\tau$ increases in the direction of \mbox{}%
\protect\/{\protect\em decreasing\protect\/} co\-ordinate time $%
t$: \setcounter{Ceqindent}{0}\protect\begin{eqnarray}\hspace{-1.%
3ex}&\displaystyle d\tau=-\mbox{$\protect\displaystyle\protect
\frac{dt}{\gamma}$}.\protect\nonumber\setlength{\Cscr}{\value{CE%
ht}\Ctenthex}\addtolength{\Cscr}{-1.0ex}\protect\raisebox{0ex}[%
\value{CEht}\Ctenthex][\Cscr]{}\protect\end{eqnarray}\setcounter
{CEht}{10}Thus, if we insist on the invariance of classical mech%
anics under the complete Lorentz group (as we do in all other fo%
rms of relativistic mechanics), we find that for every possible 
solution of the equations of motion with the choice of the \mbox
{}\protect\/{\protect\em positive\protect\/} sign in Eq.~(%
\protect\ref{eq:Tau-DTau}), there exists an equally possible sol%
ution in which the \mbox{}\protect\/{\protect\em negative\protect
\/} sign is chosen. By the Stueckelberg--Feynman interpretation, 
this transformation is the classical \mbox{}\protect\/{\protect
\em particle--antiparticle\protect\/} (or ``charge-conjugation''%
) transformation, so let us label it as such: \setcounter{Ceqind%
ent}{0}\protect\begin{eqnarray}\protect\left.\protect\begin{arra%
y}{rcl}\protect\displaystyle\hspace{-1.3ex}&\protect\displaystyle
{\cal C}:\;\tau\rightarrow-\tau.\setlength{\Cscr}{\value{CEht}%
\Ctenthex}\addtolength{\Cscr}{-1.0ex}\protect\raisebox{0ex}[%
\value{CEht}\Ctenthex][\Cscr]{}\protect\end{array}\protect\right
.\protect\label{eq:CPT-C}\protect\end{eqnarray}\setcounter{CEht}%
{10}Let us denote by $\beta$ the choice of sign in Eq.~(\protect
\ref{eq:Tau-DTau}): \mbox{$\beta=+1$} if the particle is in ``no%
rmal particle'' motion with respect to our own Lorentz frame (i.%
e., \mbox{$d\tau/dt>0$}); whereas \mbox{$\beta=-1$} if the parti%
cle is in ``antiparticle'' motion (i.e., \mbox{$d\tau/dt<0$}). (%
The symbol $\beta$ is sometimes used in introductory texts for t%
he ratio $v/c$, but in natural units it is simpler to just use t%
he intuitive symbol $v$ for this latter quantity. It is to maint%
ain consistency with the results of the Foldy--Wouthuysen transf%
ormation$^{}$$^{\ref{cit:Costella1995}}$$^{,}$$^{\ref{cit:Foldy1%
950}}${} that we use the symbol $\beta$ for the classical partic%
le--antiparticle number.)\par With the definition (\protect\ref{%
eq:CPT-C}), classical mechanics possesses all three discrete sym%
metry operations ${\cal C}$, ${\cal P}$, and ${\cal T}$ required 
for a relativistically invariant system of mechanics. Classicall%
y, all of these operations commute, and each of them individuall%
y squares to unity. We expect that the equations of motion of cl%
assical physics will be invariant under the combined operation $%
{\cal C}{\cal P}{\cal T}$.\par\refstepcounter{section}\vspace{1.%
5\baselineskip}\par{\centering\bf\thesection. Antiparticles in t%
he real world\\*[0.5\baselineskip]}\protect\indent\label{sect:Re%
alWorld}Let us now show that the classical ${\cal C}$ operation 
yields ``antiparticles'' in the everyday sense of the word, i.e.%
, that the antiparticle of an electron is a positron, and so on. 
To do so, it suffices to give our classical point particle two c%
haracteristics: an electric charge $q$, and a mass $m$. We assum%
e that the quantities $q$ and $m$ are Lorentz scalars, and are 
\mbox{}\protect\/{\protect\em unchanged\protect\/} under any of 
the operations ${\cal C}$, ${\cal P}$, or ${\cal T}$, as defined 
above. (We shall show how the usual interpretation of ${\cal C}$ 
as changing the sign of the ``effective'' charge is to be unders%
tood shortly.)\par Let us first consider the \mbox{}\protect\/{%
\protect\em electromagnetic\protect\/} interaction. This is a 
\mbox{}\protect\/{\protect\em vector\protect\/} interaction, whi%
ch couples to the \mbox{}\protect\/{\protect\em electromagnetic 
vector current density\protect\/} of the point charge,$^{\ref{ci%
t:Jackson1975}}$ \setcounter{Ceqindent}{0}\protect\begin{eqnarra%
y}\protect\left.\protect\begin{array}{rcl}\protect\displaystyle
\hspace{-1.3ex}&\protect\displaystyle j^\mu(x)\equiv q\hspace{-0%
.5mm}\protect\mbox{}\hspace{-0.1mm}\protect\int_{-\infty}^{\infty
}\protect\mbox{}\hspace{-0.5mm}\hspace{-0.6mm}d\tau\,\delta^{(4)%
}[x-z(\tau)]\,u^\mu(\tau),\setlength{\Cscr}{\value{CEht}\Ctenthex
}\addtolength{\Cscr}{-1.0ex}\protect\raisebox{0ex}[\value{CEht}%
\Ctenthex][\Cscr]{}\protect\end{array}\protect\right.\protect
\label{eq:RealWorld-J}\protect\end{eqnarray}\setcounter{CEht}{10%
}where $u^\mu(\tau)$ is the \mbox{}\protect\/{\protect\em four-v%
elocity\protect\/} of the particle: \setcounter{Ceqindent}{0}%
\protect\begin{eqnarray}\protect\left.\protect\begin{array}{rcl}%
\protect\displaystyle\hspace{-1.3ex}&\protect\displaystyle u^\mu
\equiv\mbox{$\protect\displaystyle\protect\frac{dz^\mu}{d\tau}$}%
=(\beta\gamma,\beta\gamma{\protect\mbox{\protect\boldmath{$v$}}}%
).\setlength{\Cscr}{\value{CEht}\Ctenthex}\addtolength{\Cscr}{-1%
.0ex}\protect\raisebox{0ex}[\value{CEht}\Ctenthex][\Cscr]{}%
\protect\end{array}\protect\right.\protect\label{eq:RealWorld-U}%
\protect\end{eqnarray}\setcounter{CEht}{10}The delta function an%
d $\tau$-integration in (\protect\ref{eq:RealWorld-J}) are neces%
sary to obtain a current \mbox{}\protect\/{\protect\em density%
\protect\/} from the trajectory of the point particle (the densi%
ty of a point particle being infinite on its worldline and zero 
outside); but the essential properties of $j^\mu(x)$ are contain%
ed in the classical \mbox{}\protect\/{\protect\em electromagneti%
c current vector\protect\/}, \setcounter{Ceqindent}{0}\protect
\begin{eqnarray}\protect\left.\protect\begin{array}{rcl}\protect
\displaystyle\hspace{-1.3ex}&\protect\displaystyle J^\mu\equiv q%
u^\mu.\setlength{\Cscr}{\value{CEht}\Ctenthex}\addtolength{\Cscr
}{-1.0ex}\protect\raisebox{0ex}[\value{CEht}\Ctenthex][\Cscr]{}%
\protect\end{array}\protect\right.\protect\label{eq:RealWorld-Bi%
gJ}\protect\end{eqnarray}\setcounter{CEht}{10}{}From\ Eq.~(%
\protect\ref{eq:RealWorld-U}) we see that $u^\mu$ is both ${\cal
C}$-odd and \mbox{${\cal P}{\cal T}\!$}-odd, since $dz^\mu$ is $%
{\cal C}$-even and \mbox{${\cal P}{\cal T}\!$}-odd, whereas $d%
\tau$ is by definition ${\cal C}$-odd and \mbox{${\cal P}{\cal T%
}\!$}-even. Thus, since $q$ is assumed to be unchanged by ${\cal
C}$ or ${\cal P}{\cal T}$ as we have defined them, $J^\mu$ is al%
so ${\cal C}$-odd and \mbox{${\cal P}{\cal T}\!$}-odd. (The dens%
ity $j^\mu(x)$ clearly possesses the same symmetries as $J^\mu$, 
since the delta function in (\protect\ref{eq:RealWorld-J}) is ef%
fectively an even function of its argument, and under the ${\cal
C}$ operation the sign of $d\tau$ is reversed, but so too are th%
e limits of integration.) The former property is of particular i%
mportance for us: \setcounter{Ceqindent}{0}\protect\begin{eqnarr%
ay}\protect\left.\protect\begin{array}{rcl}\protect\displaystyle
\hspace{-1.3ex}&\protect\displaystyle{\cal C}:\;J^\mu\rightarrow
-J^\mu.\setlength{\Cscr}{\value{CEht}\Ctenthex}\addtolength{\Cscr
}{-1.0ex}\protect\raisebox{0ex}[\value{CEht}\Ctenthex][\Cscr]{}%
\protect\end{array}\protect\right.\protect\label{eq:RealWorld-CJ%
}\protect\end{eqnarray}\setcounter{CEht}{10}The result (\protect
\ref{eq:RealWorld-CJ}) tells us that, as far the electromagnetic 
interaction is concerned, antiparticle (\mbox{$\beta=-1$}) motio%
n of a charged particle \mbox{}\protect\/{\protect\em appears th%
e same\protect\/} as an ``equivalent normal particle'', with 
\mbox{$\beta=+1$}, with the same three-velocity ${\protect\mbox{%
\protect\boldmath{$v$}}}$ as the original particle, but with the 
\mbox{}\protect\/{\protect\em opposite ``effective'' charge%
\protect\/}, since \setcounter{Ceqindent}{0}\protect\begin{eqnar%
ray}\hspace{-1.3ex}&\displaystyle J^\mu\equiv q\,\mbox{$\protect
\displaystyle\protect\frac{dz^\mu}{d\tau}$}=(-q)\,\mbox{$\protect
\displaystyle\protect\frac{dz^\mu}{d(-\tau)}$}.\protect\nonumber
\setlength{\Cscr}{\value{CEht}\Ctenthex}\addtolength{\Cscr}{-1.0%
ex}\protect\raisebox{0ex}[\value{CEht}\Ctenthex][\Cscr]{}\protect
\end{eqnarray}\setcounter{CEht}{10}For example, a particle of ch%
arge $q$, at rest, but in antiparticle motion, generates a stati%
c Coulomb field that is equivalent to that from a ``normal'' par%
ticle at rest of charge $-q$. It is in this sense---of replacing 
antiparticle motion by an ``equivalent normal particle'' with an 
opposite ``effective'' charge---that the classical particle--ant%
iparticle operation ${\cal C}$ is a ``charge conjugation'' opera%
tion.\par Let us now determine the \mbox{}\protect\/{\protect\em
mass\protect\/} of the ``equivalent normal particle'' correspond%
ing to antiparticle motion. A~priori, it may not be clear how we 
can make such a determination. However, if we believe Einstein's 
theory of general relativity (which we do here), then the equiva%
lence principle tells us that ``inertial'' mass and ``gravitatio%
nal'' mass are one and the same thing: they both represent the c%
oupling of \mbox{}\protect\/{\protect\em matter\protect\/} to 
\mbox{}\protect\/{\protect\em spacetime\protect\/}. Thus, we can 
simply carry out the same analysis as we performed above for the 
electromagnetic interaction, but now with regard to the \mbox{}%
\protect\/{\protect\em gravitational\protect\/} interaction.\par
Gravitation is a \mbox{}\protect\/{\protect\em tensor\protect\/} 
interaction, which couples to the \mbox{}\protect\/{\protect\em
mechanical stress--energy tensor current density\protect\/} of o%
ur point particle: \setcounter{Ceqindent}{0}\protect\begin{eqnar%
ray}\protect\left.\protect\begin{array}{rcl}\protect\displaystyle
\hspace{-1.3ex}&\protect\displaystyle t^{\mu\nu}(x)\equiv m%
\hspace{-0.5mm}\protect\mbox{}\hspace{-0.1mm}\protect\int_{-%
\infty}^{\infty}\protect\mbox{}\hspace{-0.5mm}\hspace{-0.6mm}d%
\tau\,\delta^{(4)}[x-z(\tau)]\,u^\mu(\tau)u^\nu(\tau).\setlength
{\Cscr}{\value{CEht}\Ctenthex}\addtolength{\Cscr}{-1.0ex}\protect
\raisebox{0ex}[\value{CEht}\Ctenthex][\Cscr]{}\protect\end{array%
}\protect\right.\protect\label{eq:RealWorld-TDensity}\protect\end
{eqnarray}\setcounter{CEht}{10}Again, the delta function and int%
egration over $\tau$ and are required for the purposes of conver%
ting a pointlike trajectory into a \mbox{}\protect\/{\protect\em
density\protect\/}; the essential properties of $t^{\mu\nu}(x)$ 
are contained in the classical \mbox{}\protect\/{\protect\em mec%
hanical current tensor\protect\/}, \setcounter{Ceqindent}{0}%
\protect\begin{eqnarray}\protect\left.\protect\begin{array}{rcl}%
\protect\displaystyle\hspace{-1.3ex}&\protect\displaystyle T^{\mu
\nu}\equiv mu^\mu u^\nu.\setlength{\Cscr}{\value{CEht}\Ctenthex}%
\addtolength{\Cscr}{-1.0ex}\protect\raisebox{0ex}[\value{CEht}%
\Ctenthex][\Cscr]{}\protect\end{array}\protect\right.\protect
\label{eq:RealWorld-BigT}\protect\end{eqnarray}\setcounter{CEht}%
{10}In this case, we find that $T^{\mu\nu}$ is ${\cal C}$-\mbox{%
}\protect\/{\protect\em even\protect\/}, since it contains \mbox
{}\protect\/{\protect\em two\protect\/} factors of $u^\mu$: 
\setcounter{Ceqindent}{0}\protect\begin{eqnarray}\hspace{-1.3ex}%
&\displaystyle{\cal C}:\;T^{\mu\nu}\rightarrow T^{\mu\nu}.%
\protect\nonumber\setlength{\Cscr}{\value{CEht}\Ctenthex}%
\addtolength{\Cscr}{-1.0ex}\protect\raisebox{0ex}[\value{CEht}%
\Ctenthex][\Cscr]{}\protect\end{eqnarray}\setcounter{CEht}{10}Th%
us, as far as the gravitational interaction is concerned, a part%
icle of mass $m$ in antiparticle motion (\mbox{$\beta=-1$}) is i%
ndistinguishable from the same particle of mass $m$ in the corre%
sponding particle motion (\mbox{$\beta=+1$}). For example, the g%
ravitational field of a star made of antimatter is the \mbox{}%
\protect\/{\protect\em same\protect\/} as that of an identical s%
tar in which the antimatter is replaced by normal matter; a coll%
ection of such stars would all \mbox{}\protect\/{\protect\em att%
ract\protect\/} each other gravitationally. By the equivalence p%
rinciple, this invariance of the mass of the ``equivalent normal 
particle'' under the ${\cal C}$ operation is true in full genera%
lity.\par The above examples have concentrated on the fields 
\mbox{}\protect\/{\protect\em generated\protect\/} by particles 
in antiparticle motion, but the same conclusions can be drawn fr%
om the equations of motion for the particles \mbox{}\protect\/{%
\protect\em under the influence\protect\/} of given external fie%
lds. In the gravitational case, the mass $m$ actually drops out 
of the equations of motion (again, by the equivalence principle)%
, and the equation of motion is simply the \mbox{}\protect\/{%
\protect\em geodesic equation\protect\/},$^{\ref{cit:Misner1970}%
}$ \setcounter{Ceqindent}{0}\protect\begin{eqnarray}\protect\left
.\protect\begin{array}{rcl}\protect\displaystyle\hspace{-1.3ex}&%
\protect\displaystyle\mbox{$\protect\displaystyle\protect\frac{d%
u^\mu}{d\tau}$}=-{{\protect\it\Gamma\!\:}^\mu}_{\!\!\nu\rho\,}u^%
\nu u^\rho.\setlength{\Cscr}{\value{CEht}\Ctenthex}\addtolength{%
\Cscr}{-1.0ex}\protect\raisebox{0ex}[\value{CEht}\Ctenthex][\Cscr
]{}\protect\end{array}\protect\right.\protect\label{eq:RealWorld%
-Geodesic}\protect\end{eqnarray}\setcounter{CEht}{10}If the part%
icle is in antiparticle motion, we can again replace it by an ``%
equivalent normal particle'': since \setcounter{Ceqindent}{0}%
\protect\begin{eqnarray}\protect\left.\protect\begin{array}{rcl}%
\protect\displaystyle\hspace{-1.3ex}&\protect\displaystyle\mbox{%
$\protect\displaystyle\protect\frac{du^\mu}{d\tau}$}\equiv\mbox{%
$\protect\displaystyle\protect\frac{d^{\hspace{0.15ex}2}\hspace{%
-0.45mm}z^\mu}{d\tau^2}$}=\mbox{$\protect\displaystyle\protect
\frac{d^{\hspace{0.15ex}2}\hspace{-0.45mm}z^\mu}{d(-\tau)^2}$}%
\setlength{\Cscr}{\value{CEht}\Ctenthex}\addtolength{\Cscr}{-1.0%
ex}\protect\raisebox{0ex}[\value{CEht}\Ctenthex][\Cscr]{}\protect
\end{array}\protect\right.\protect\label{eq:RealWorld-SecondDeri%
v}\protect\end{eqnarray}\setcounter{CEht}{10}and \setcounter{Ceq%
indent}{0}\protect\begin{eqnarray}\protect\left.\protect\begin{a%
rray}{rcl}\protect\displaystyle\hspace{-1.3ex}&\protect
\displaystyle u^\mu\equiv\mbox{$\protect\displaystyle\protect
\frac{dz^\mu}{d\tau}$}=-\mbox{$\protect\displaystyle\protect\frac
{dz^\mu}{d(-\tau)}$},\setlength{\Cscr}{\value{CEht}\Ctenthex}%
\addtolength{\Cscr}{-1.0ex}\protect\raisebox{0ex}[\value{CEht}%
\Ctenthex][\Cscr]{}\protect\end{array}\protect\right.\protect
\label{eq:RealWorld-FirstDeriv}\protect\end{eqnarray}\setcounter
{CEht}{10}we can rewrite the geodesic equation (\protect\ref{eq:%
RealWorld-Geodesic}) in the form \setcounter{Ceqindent}{0}%
\protect\begin{eqnarray}\hspace{-1.3ex}&\displaystyle\mbox{$%
\protect\displaystyle\protect\frac{d^{\hspace{0.15ex}2}\hspace{-%
0.45mm}z^\mu}{d(-\tau)^2}$}=-{{\protect\it\Gamma\!\:}^\mu}_{\!\!%
\nu\rho\,}\mbox{$\protect\displaystyle\protect\frac{dz^\nu}{d(-%
\tau)}$}\mbox{$\protect\displaystyle\protect\frac{dz^\rho}{d(-%
\tau)}$},\protect\nonumber\setlength{\Cscr}{\value{CEht}\Ctenthex
}\addtolength{\Cscr}{-1.0ex}\protect\raisebox{0ex}[\value{CEht}%
\Ctenthex][\Cscr]{}\protect\end{eqnarray}\setcounter{CEht}{10}wh%
ich shows that the ``equivalent normal particle'' acts in the sa%
me way as any other particle.\par In the electromagnetic case, t%
he equation of motion is the \mbox{}\protect\/{\protect\em Loren%
tz force law\protect\/}, which in relativistic form is 
\setcounter{Ceqindent}{0}\protect\begin{eqnarray}\protect\left.%
\protect\begin{array}{rcl}\protect\displaystyle\hspace{-1.3ex}&%
\protect\displaystyle\mbox{$\protect\displaystyle\protect\frac{d%
u^\mu}{d\tau}$}=\mbox{$\protect\displaystyle\protect\frac{q}{m}$%
}F^{\mu\nu\!}u_\nu,\setlength{\Cscr}{\value{CEht}\Ctenthex}%
\addtolength{\Cscr}{-1.0ex}\protect\raisebox{0ex}[\value{CEht}%
\Ctenthex][\Cscr]{}\protect\end{array}\protect\right.\protect
\label{eq:RealWorld-LorentzForce}\protect\end{eqnarray}%
\setcounter{CEht}{10}where $F^{\mu\nu}$ represents an external e%
lectromagnetic field. Again, if the particle is in antiparticle 
motion, we can use the properties (\protect\ref{eq:RealWorld-Sec%
ondDeriv}) and (\protect\ref{eq:RealWorld-FirstDeriv}) to rewrit%
e Eq.~(\protect\ref{eq:RealWorld-LorentzForce}) in the form 
\setcounter{Ceqindent}{0}\protect\begin{eqnarray}\hspace{-1.3ex}%
&\displaystyle\mbox{$\protect\displaystyle\protect\frac{d^{%
\hspace{0.15ex}2}\hspace{-0.45mm}z^\mu}{d(-\tau)^2}$}=-\mbox{$%
\protect\displaystyle\protect\frac{q}{m}$}F^{\mu\nu\!}\mbox{$%
\protect\displaystyle\protect\frac{dz_\nu}{d(-\tau)}$}.\protect
\nonumber\setlength{\Cscr}{\value{CEht}\Ctenthex}\addtolength{%
\Cscr}{-1.0ex}\protect\raisebox{0ex}[\value{CEht}\Ctenthex][\Cscr
]{}\protect\end{eqnarray}\setcounter{CEht}{10}Hence the ``equiva%
lent normal particle'' has the sign of its effective \mbox{}%
\protect\/{\protect\em charge-to-mass ratio\protect\/} negated c%
ompared to that of the actual particle; and since we have alread%
y determined that its \mbox{}\protect\/{\protect\em mass\protect
\/} is unchanged, we conclude that it is its ``effective charge%
'' that changes sign, in agreement with our analysis above.\par
\refstepcounter{section}\vspace{1.5\baselineskip}\par{\centering
\bf\thesection. The ``negative energy problem''\\*[0.5%
\baselineskip]}\protect\indent\label{sect:NegEn}Finally, let us 
discuss a subtlety that is a frequent source of confusion for th%
e student: the existence and interpretation of an \mbox{}\protect
\/{\protect\em energy--momentum four-vector\protect\/} for antip%
articles. The subject arises most naturally when we consider the 
Lagrangian and Hamiltonian formulations of mechanics (which are 
of course of vital importance in the construction of a quantum m%
echanical description); but in introductory courses and textbook%
s it is often presented in a somewhat confusing and contradictor%
y manner.\par Let us continue to consider our classical point pa%
rticle of mass $m$ and charge $q$. Relativistically, the constru%
ction of a manifestly covariant set of generalized co\-ordinates 
is somewhat delicate:$^{}$$^{\ref{cit:Jackson1975}}$$^{,}$$^{\ref
{cit:Goldstein1980}}${} we would like to treat \mbox{}\protect\/%
{\protect\em all four\protect\/} components $z^\mu$ of the posit%
ion four-vector of the particle in an equal fashion, with the ``%
time'' parameter preferentially given by the Lorentz-invariant p%
roper time $\tau$. But the existence of the definition (\protect
\ref{eq:Tau-DTauSquared}) tells us that only \mbox{}\protect\/{%
\protect\em three\protect\/} of the components of $z^\mu$ are ac%
tually independent of $\tau$; and the Lagrangian and Hamiltonian 
formulations of mechanics are greatly complicated if all of the 
generalized co\-ordinates are not actually independent.$^{\ref{c%
it:Goldstein1980}}$\par One way to circumvent these problems is 
to take the ``time'' parameter of the Lagrangian or Hamiltonian 
formalism to be the ``$\theta$-time'', where \setcounter{Ceqinde%
nt}{0}\protect\begin{eqnarray}\protect\left.\protect\begin{array%
}{rcl}\protect\displaystyle\hspace{-1.3ex}&\protect\displaystyle
d\theta\equiv\mbox{$\protect\displaystyle\protect\frac{d\tau}{m}%
$}.\setlength{\Cscr}{\value{CEht}\Ctenthex}\addtolength{\Cscr}{-%
1.0ex}\protect\raisebox{0ex}[\value{CEht}\Ctenthex][\Cscr]{}%
\protect\end{array}\protect\right.\protect\label{eq:NegEn-Theta}%
\protect\end{eqnarray}\setcounter{CEht}{10}The quantity $\theta$ 
is a Lorentz scalar, since both $\tau$ and $m$ are Lorentz scala%
rs; and since $m$ is even under ${\cal C}$, ${\cal P}$, and ${%
\cal T}$, the quantity $\theta$ has the same symmetry properties 
as $\tau$; in particular, \setcounter{Ceqindent}{0}\protect\begin
{eqnarray}\hspace{-1.3ex}&\displaystyle{\cal C}:\;\theta
\rightarrow-\theta.\protect\nonumber\setlength{\Cscr}{\value{CEh%
t}\Ctenthex}\addtolength{\Cscr}{-1.0ex}\protect\raisebox{0ex}[%
\value{CEht}\Ctenthex][\Cscr]{}\protect\end{eqnarray}\setcounter
{CEht}{10}To show that the simple redefinition (\protect\ref{eq:%
NegEn-Theta}) solves the independence problem, one need simply n%
ote that the \mbox{}\protect\/{\protect\em mass\protect\/} (rest%
-energy) $m(\tau)$ of a \mbox{}\protect\/{\protect\em general%
\protect\/} system need not be a constant of the motion. Thus, w%
hile one of the four components of $z^\mu$ is still differential%
ly dependent on $\tau$ through Eq.~(\protect\ref{eq:Tau-DTauSqua%
red}), \mbox{}\protect\/{\protect\em all four\protect\/} compone%
nts of $z^\mu$ are, in general, independent of $\theta$, since t%
he $m$ in Eq.~(\protect\ref{eq:NegEn-Theta}) may vary as a funct%
ion of $\tau$. A useful bonus of this approach is that \mbox{}%
\protect\/{\protect\em all\protect\/} equations in the $\theta$-%
time formalism can be written in such a way that that the quanti%
ties $m$ and $\tau$ never explicitly appear; the formalism may t%
hen be applied equally well to \mbox{}\protect\/{\protect\em mas%
sless\protect\/} particles (for which $\theta$ exists, but for w%
hich \mbox{$m=0$} and $\tau$ is undefinable).\par A somewhat sim%
plified version of this approach is not to actually use the $%
\theta$-time at all, but rather to simply ``pretend'' that $\tau
$ is not in fact constrained by Eq.~(\protect\ref{eq:Tau-DTauSqu%
ared}) until \mbox{}\protect\/{\protect\em after\protect\/} the 
equations of motion have been obtained. The results are the same%
, so let us follow this latter, simplified approach. A suitable 
Lagrangian can, for example, be chosen to be$^{\ref{cit:Stueckel%
berg1942}}$ \setcounter{Ceqindent}{0}\protect\begin{eqnarray}%
\protect\left.\protect\begin{array}{rcl}\protect\displaystyle
\hspace{-1.3ex}&\protect\displaystyle L=\mbox{$\protect
\displaystyle\protect\frac{1}{2}$}mu^\mu u_\mu+qu^\mu\!A_\mu,%
\setlength{\Cscr}{\value{CEht}\Ctenthex}\addtolength{\Cscr}{-1.0%
ex}\protect\raisebox{0ex}[\value{CEht}\Ctenthex][\Cscr]{}\protect
\end{array}\protect\right.\protect\label{eq:NegEn-L}\protect\end
{eqnarray}\setcounter{CEht}{10}where $A_\mu$ is the electromagne%
tic four-potential: \setcounter{Ceqindent}{0}\protect\begin{eqna%
rray}\hspace{-1.3ex}&\displaystyle A^\mu\equiv(\varphi,{\protect
\mbox{\protect\boldmath{$A$}}}).\protect\nonumber\setlength{\Cscr
}{\value{CEht}\Ctenthex}\addtolength{\Cscr}{-1.0ex}\protect
\raisebox{0ex}[\value{CEht}\Ctenthex][\Cscr]{}\protect\end{eqnar%
ray}\setcounter{CEht}{10}Using Eqs.~(\protect\ref{eq:RealWorld-B%
igJ}) and (\protect\ref{eq:RealWorld-BigT}), we can recognize th%
e Lagrangian (\protect\ref{eq:NegEn-L}) as simply a straightforw%
ard coupling of the tensor and vector currents of the particle t%
o their corresponding gauge fields: \setcounter{Ceqindent}{0}%
\protect\begin{eqnarray}\hspace{-1.3ex}&\displaystyle L=\mbox{$%
\protect\displaystyle\protect\frac{1}{2}$}T^{\mu\nu\!}g_{\mu\nu}%
+J^\mu\!A_\mu.\protect\nonumber\setlength{\Cscr}{\value{CEht}%
\Ctenthex}\addtolength{\Cscr}{-1.0ex}\protect\raisebox{0ex}[%
\value{CEht}\Ctenthex][\Cscr]{}\protect\end{eqnarray}\setcounter
{CEht}{10}The canonical momentum components $p_\mu$ conjugate to 
the generalized degrees of freedom $z^\mu$ are obtained from (%
\protect\ref{eq:NegEn-L}) in the standard way: \setcounter{Ceqin%
dent}{0}\protect\begin{eqnarray}\protect\left.\protect\begin{arr%
ay}{rcl}\protect\displaystyle\hspace{-1.3ex}&\protect
\displaystyle p_\mu\equiv\mbox{$\protect\displaystyle\protect
\frac{\partial L}{\partial u^\mu}$}=mu_\mu+qA_\mu.\setlength{%
\Cscr}{\value{CEht}\Ctenthex}\addtolength{\Cscr}{-1.0ex}\protect
\raisebox{0ex}[\value{CEht}\Ctenthex][\Cscr]{}\protect\end{array%
}\protect\right.\protect\label{eq:NegEn-CanMom}\protect\end{eqna%
rray}\setcounter{CEht}{10}The Euler--Lagrange equations of motio%
n then yield the Lorentz force law (\protect\ref{eq:RealWorld-Lo%
rentzForce}).$^{\ref{cit:Costella1994}}$\par A corresponding 
\mbox{}\protect\/{\protect\em Hamiltonian\protect\/} formulation 
of this same classical theory may be constructed in two differen%
t ways. On the one hand, a ``manifestly-covariant Hamiltonian'' 
${\cal H}$ can be constructed by the usual Legendre transformati%
on: \setcounter{Ceqindent}{0}\protect\begin{eqnarray}\hspace{-1.%
3ex}&\displaystyle{\cal H}\equiv p_\mu u^\mu-L=\mbox{$\protect
\displaystyle\protect\frac{1}{2m}$}(p^\mu-qA^\mu)(p_\mu-qA_\mu)%
\equiv\mbox{$\protect\displaystyle\protect\frac{(p-qA)^2}{2m}$},%
\protect\nonumber\setlength{\Cscr}{\value{CEht}\Ctenthex}%
\addtolength{\Cscr}{-1.0ex}\protect\raisebox{0ex}[\value{CEht}%
\Ctenthex][\Cscr]{}\protect\end{eqnarray}\setcounter{CEht}{10}fr%
om which Hamilton's equations (with respect to the ``time'' para%
meter $\tau$) again yield the Lorentz force law (\protect\ref{eq%
:RealWorld-LorentzForce}), as well as the equation (\protect\ref
{eq:NegEn-CanMom}) relating the canonical momentum and velocity 
four-vectors.\par On the other hand, the more \mbox{}\protect\/{%
\protect\em conventional\protect\/} way to construct a Hamiltoni%
an formulation of this system---that yields a somewhat simpler t%
ransition to the quantum theory---is to recognize that the \mbox
{}\protect\/{\protect\em canonical energy\protect\/} $p^0$ can b%
e interpreted as the Hamiltonian $H$ of the system, with respect 
to the co\-ordinate time $t$. {}From\ Eq.~(\protect\ref{eq:NegEn%
-CanMom}), we have \setcounter{Ceqindent}{0}\protect\begin{eqnar%
ray}\hspace{-1.3ex}&\displaystyle(p-qA)^2=m^2(u^\mu u_\mu)\equiv
m^2,\protect\nonumber\setlength{\Cscr}{\value{CEht}\Ctenthex}%
\addtolength{\Cscr}{-1.0ex}\protect\raisebox{0ex}[\value{CEht}%
\Ctenthex][\Cscr]{}\protect\end{eqnarray}\setcounter{CEht}{10}wh%
ere in this last expression we can ``stop pretending'' that $u^%
\mu u_\mu$ is not identically equal to unity, because it is $p^%
\mu$, not $u^\mu$, that plays a fundamental role\ in the Hamilto%
nian formulation of mechanics. Thus \setcounter{Ceqindent}{0}%
\protect\begin{eqnarray}\hspace{-1.3ex}&\displaystyle(H-q\varphi
)^2-({\protect\mbox{\protect\boldmath{$p$}}}-q{\protect\mbox{%
\protect\boldmath{$A$}}})^2=m^2,\protect\nonumber\setlength{\Cscr
}{\value{CEht}\Ctenthex}\addtolength{\Cscr}{-1.0ex}\protect
\raisebox{0ex}[\value{CEht}\Ctenthex][\Cscr]{}\protect\end{eqnar%
ray}\setcounter{CEht}{10}and hence \setcounter{Ceqindent}{0}%
\protect\begin{eqnarray}\protect\left.\protect\begin{array}{rcl}%
\protect\displaystyle H=q\varphi+\beta\sqrt{m^2+({\protect\mbox{%
\protect\boldmath{$p$}}}-q{\protect\mbox{\protect\boldmath{$A$}}%
})^2},\setlength{\Cscr}{\value{CEht}\Ctenthex}\addtolength{\Cscr
}{-1.0ex}\protect\raisebox{0ex}[\value{CEht}\Ctenthex][\Cscr]{}%
\protect\end{array}\protect\right.\protect\label{eq:NegEn-NonCov%
H}\protect\end{eqnarray}\setcounter{CEht}{10}where \mbox{$\beta=%
\pm1$} encapsulates the choice of sign in taking the square root%
. Hamilton's equations again yield expressions equivalent to (%
\protect\ref{eq:RealWorld-LorentzForce}) and (\protect\ref{eq:Ne%
gEn-CanMom}).\par Let us now consider the simple case of a free 
particle, which will be sufficient for our purposes. {}From\ Eq.%
 ~(\protect\ref{eq:NegEn-CanMom}), we have in this case 
\setcounter{Ceqindent}{0}\protect\begin{eqnarray}\protect\left.%
\protect\begin{array}{rcl}\protect\displaystyle\hspace{-1.3ex}&%
\protect\displaystyle p^\mu=mu^\mu;\setlength{\Cscr}{\value{CEht%
}\Ctenthex}\addtolength{\Cscr}{-1.0ex}\protect\raisebox{0ex}[%
\value{CEht}\Ctenthex][\Cscr]{}\protect\end{array}\protect\right
.\protect\label{eq:NegEn-PFree}\protect\end{eqnarray}\setcounter
{CEht}{10}equivalently, from Eq.~(\protect\ref{eq:NegEn-NonCovH}%
), we have \setcounter{Ceqindent}{0}\protect\begin{eqnarray}%
\protect\left.\protect\begin{array}{rcl}\protect\displaystyle
\hspace{-1.3ex}&\protect\displaystyle H=\beta\sqrt{m^2+{\protect
\mbox{\protect\boldmath{$p$}}}^2}.\setlength{\Cscr}{\value{CEht}%
\Ctenthex}\addtolength{\Cscr}{-1.0ex}\protect\raisebox{0ex}[%
\value{CEht}\Ctenthex][\Cscr]{}\protect\end{array}\protect\right
.\protect\label{eq:NegEn-HFree}\protect\end{eqnarray}\setcounter
{CEht}{10}Eqs.~(\protect\ref{eq:NegEn-PFree}) and (\protect\ref{%
eq:NegEn-HFree}) emphasize the fact that, in antiparticle motion%
, the canonical momentum four-vector $p^\mu$ has \mbox{}\protect
\/{\protect\em negative energy\protect\/}; or, in other words, t%
hat $p^\mu$ is \mbox{}\protect\/{\protect\em odd\protect\/} unde%
r ${\cal C}$: \setcounter{Ceqindent}{0}\protect\begin{eqnarray}%
\hspace{-1.3ex}&\displaystyle{\cal C}:\;p^\mu\rightarrow-p^\mu.%
\protect\nonumber\setlength{\Cscr}{\value{CEht}\Ctenthex}%
\addtolength{\Cscr}{-1.0ex}\protect\raisebox{0ex}[\value{CEht}%
\Ctenthex][\Cscr]{}\protect\end{eqnarray}\setcounter{CEht}{10}%
\par Consider, now, a particle and its corresponding antiparticl%
e, both at rest. For the former, we have \setcounter{Ceqindent}{%
0}\protect\begin{eqnarray}\protect\left.\protect\begin{array}{rc%
l}\protect\displaystyle\hspace{-1.3ex}&\protect\displaystyle p^%
\mu=(m,\mbox{\bf0}),\setlength{\Cscr}{\value{CEht}\Ctenthex}%
\addtolength{\Cscr}{-1.0ex}\protect\raisebox{0ex}[\value{CEht}%
\Ctenthex][\Cscr]{}\protect\end{array}\protect\right.\protect
\label{eq:NegEn-ElectronP}\protect\end{eqnarray}\setcounter{CEht%
}{10}whereas for the latter we have \setcounter{Ceqindent}{0}%
\protect\begin{eqnarray}\protect\left.\protect\begin{array}{rcl}%
\protect\displaystyle\hspace{-1.3ex}&\protect\displaystyle p^\mu
=(-m,\mbox{\bf0}).\setlength{\Cscr}{\value{CEht}\Ctenthex}%
\addtolength{\Cscr}{-1.0ex}\protect\raisebox{0ex}[\value{CEht}%
\Ctenthex][\Cscr]{}\protect\end{array}\protect\right.\protect
\label{eq:NegEn-PositronP}\protect\end{eqnarray}\setcounter{CEht%
}{10}``Surely,'' a common argument goes, ``does this not tell us 
that antiparticle motion is really a \mbox{}\protect\/{\protect
\em negative mass\protect\/} solution? Does it not further tell 
us that the \mbox{}\protect\/{\protect\em total\protect\/} energ%
y of this pair of particles is \mbox{}\protect\/{\protect\em zer%
o\protect\/}?'' These two statements directly contradict our fin%
ding above that the mass of a particle is \mbox{}\protect\/{%
\protect\em unchanged\protect\/}, whether it be in particle or a%
ntiparticle motion; and we of course know that the total energy 
of a neutral particle--antiparticle pair at rest is indeed $2m$, 
\mbox{}\protect\/{\protect\em not\protect\/} zero. On the other 
hand, we know that, quantum mechanically, the canonical energy 
\mbox{}\protect\/{\protect\em really is\protect\/} negative for 
an antiparticle solution: for example, the eigenstates for \mbox
{${\protect\mbox{\protect\boldmath{$p$}}}=\mbox{\bf0}$} have a t%
ime-dependence of the form $e^{\mp imt}$ (where we use units in 
which \mbox{$\hbar=1$}), which, using the Einstein relation \mbox
{$E=i\partial/\partial t$}, implies that \mbox{$E=\pm m$}. So wh%
at are we to believe?\par This ``paradox'' is usually presented 
in discussions of relativistic quantum mechanics---leading to Di%
rac's ``holes'', and so on---but it is fundamentally a feature o%
f relativistic mechanics itself, whether it be of the classical 
\mbox{}\protect\/{\protect\em or\protect\/} quantum flavor. Let 
us now dispense with it once and for all.\par The fallacy above 
is the assumption that the canonical momentum four-vector has an%
ything at all to do with the ``total mass'' of a system. It does 
not. In trying to ``add together'' the four-momenta of the two p%
articles, we are making the implicit assumption that we are in s%
ome sense computing the four-momentum generalization of the ``to%
tal energy'' or ``total rest mass'' of a system. But we have alr%
eady noted that for a system of \mbox{}\protect\/{\protect\em gr%
avitational\protect\/} sources, it is the sum of the \mbox{}%
\protect\/{\protect\em mechanical stress-energy tensor densities%
\protect\/} $t^{\mu\nu}(x)$ that determines the overall gravitat%
ional field generated by the system. {}From\ this gravitational 
field, one can define a ``total gravitational mass'' of the syst%
em (because for our purposes the strengths of these gravitationa%
l fields are assumed to be negligible compared to the other forc%
es present, so that special relativity is a good approximation). 
But by the equivalence principle, this ``gravitational'' mass is 
simply \mbox{}\protect\/{\protect\em the\protect\/} mass of the 
system; and \mbox{}\protect\/{\protect\em no definition of ``mas%
s'' giving a different result can be compatible with the equival%
ence principle\protect\/}. Now, we know that $T^{\mu\nu}$ is 
\mbox{}\protect\/{\protect\em unchanged\protect\/} under the ${%
\cal C}$-operation; in particular, for both the particle \mbox{}%
\protect\/{\protect\em and\protect\/} the antiparticle above, we 
have \setcounter{Ceqindent}{0}\protect\begin{eqnarray}\hspace{-1%
.3ex}&\displaystyle T^{00}=m,\protect\nonumber\setlength{\Cscr}{%
\value{CEht}\Ctenthex}\addtolength{\Cscr}{-1.0ex}\protect
\raisebox{0ex}[\value{CEht}\Ctenthex][\Cscr]{}\\[0ex]\protect
\displaystyle\hspace{-1.3ex}&\displaystyle T^{0i}=T^{i0}=T^{ij}=%
0,\protect\nonumber\setlength{\Cscr}{\value{CEht}\Ctenthex}%
\addtolength{\Cscr}{-1.0ex}\protect\raisebox{0ex}[\value{CEht}%
\Ctenthex][\Cscr]{}\protect\end{eqnarray}\setcounter{CEht}{10}an%
d so for the system as a whole we have \mbox{$T^{00}=2m$}. Thus, 
the total mass of the particle--antiparticle pair at rest \mbox{%
}\protect\/{\protect\em is\protect\/} indeed the commonsense res%
ult of $2m$, not zero.\par Let us review these issues in more de%
tail. The Hamiltonian $H$ is the zero-component of the canonical 
momentum four-vector $p^\mu$, and is thus reasonably called the 
``canonical energy''. In quantum mechanics, the canonical moment%
um four-vector relates directly to frequency and wavelength: 
\setcounter{Ceqindent}{0}\protect\begin{eqnarray}p_\mu\rightarrow
i\partial_\mu.\protect\nonumber\setlength{\Cscr}{\value{CEht}%
\Ctenthex}\addtolength{\Cscr}{-1.0ex}\protect\raisebox{0ex}[%
\value{CEht}\Ctenthex][\Cscr]{}\protect\end{eqnarray}\setcounter
{CEht}{10}Thus, particle (antiparticle) motion for a free partic%
le corresponds to positive (negative) frequencies. But this has 
nothing at all to do with those \mbox{}\protect\/{\protect\em me%
chanical\protect\/} (or ``kinematical'') properties of the parti%
cle, that are physically observable. Indeed, in the presence of 
interactions, even the \mbox{}\protect\/{\protect\em sign\protect
\/} of the canonical energy (frequency) loses its relevance comp%
letely: for example, for the electromagnetic interaction of a cl%
assical point charge, we have \setcounter{Ceqindent}{0}\protect
\begin{eqnarray}p^\mu=mu^\mu+qA^\mu.\protect\nonumber\setlength{%
\Cscr}{\value{CEht}\Ctenthex}\addtolength{\Cscr}{-1.0ex}\protect
\raisebox{0ex}[\value{CEht}\Ctenthex][\Cscr]{}\protect\end{eqnar%
ray}\setcounter{CEht}{10}It will be immediately noted that $p^\mu
$ is not gauge-invariant, and the value of $p^0$ can be given an%
y arbitrary value (positive \mbox{}\protect\/{\protect\em or%
\protect\/} negative) simply by redefining the zero point of the 
scalar potential. Thus $p^\mu$ cannot possibly, of itself, deter%
mine any physically observable property of the particle, such as 
its mass, or its mechanical stress--energy tensor.\par On the ot%
her hand, we know that it \mbox{}\protect\/{\protect\em is%
\protect\/} possible to define some sort of four-vector $\pi^\mu
$ that represents the \mbox{}\protect\/{\protect\em mechanical%
\protect\/} energy--momentum of a system---after all, we have be%
en adding energies and momenta together for centuries. To define 
such a $\pi^\mu$, we need simply integrate the \mbox{}\protect\/%
{\protect\em mechanical\protect\/} stress-energy tensor density 
$t^{\mu\nu}(x)$ over all three-space, in some \mbox{}\protect\/{%
\protect\em given\protect\/} Lorentz frame. We can write this co%
variantly in the form \setcounter{Ceqindent}{0}\protect\begin{eq%
narray}\protect\left.\protect\begin{array}{rcl}\protect
\displaystyle\hspace{-1.3ex}&\protect\displaystyle\pi^\mu(t)%
\equiv\hspace{-0.5mm}\protect\mbox{}\hspace{-0.1mm}\protect\int
\protect\mbox{}\hspace{-0.5mm}\hspace{-0.6mm}d^{\hspace{0.15ex}3%
}\hspace{-0.45mm}\sigma_\nu\,t^{\mu\nu}(t,{\protect\mbox{\protect
\boldmath{$x$}}}),\setlength{\Cscr}{\value{CEht}\Ctenthex}%
\addtolength{\Cscr}{-1.0ex}\protect\raisebox{0ex}[\value{CEht}%
\Ctenthex][\Cscr]{}\protect\end{array}\protect\right.\protect
\label{eq:NegEn-DefnPi}\protect\end{eqnarray}\setcounter{CEht}{1%
0}where an element of ``three-space'' has been written covariant%
ly as $d^{\hspace{0.15ex}3}\hspace{-0.45mm}\sigma^\mu$: in the g%
iven Lorentz frame, in which it is simply all three-space at a c%
onstant time $t$, we have \setcounter{Ceqindent}{0}\protect\begin
{eqnarray}\hspace{-1.3ex}&\displaystyle d^{\hspace{0.15ex}3}%
\hspace{-0.45mm}\sigma^\mu=n^\mu\,d^{\hspace{0.15ex}3}\hspace{-0%
.45mm}x,\protect\nonumber\setlength{\Cscr}{\value{CEht}\Ctenthex
}\addtolength{\Cscr}{-1.0ex}\protect\raisebox{0ex}[\value{CEht}%
\Ctenthex][\Cscr]{}\protect\end{eqnarray}\setcounter{CEht}{10}wh%
ere $n^\mu$ is the timelike \mbox{}\protect\/{\protect\em four-n%
ormal\protect\/} to the hyperplane, which in this frame has co\-%
ordinates \setcounter{Ceqindent}{0}\protect\begin{eqnarray}%
\hspace{-1.3ex}&\displaystyle n^\mu=(1,\mbox{\bf0}).\protect
\nonumber\setlength{\Cscr}{\value{CEht}\Ctenthex}\addtolength{%
\Cscr}{-1.0ex}\protect\raisebox{0ex}[\value{CEht}\Ctenthex][\Cscr
]{}\protect\end{eqnarray}\setcounter{CEht}{10}Let us compute $\pi
^\mu$ in this Lorentz frame: from Eqs.~(\protect\ref{eq:RealWorl%
d-TDensity}) and (\protect\ref{eq:NegEn-DefnPi}), we have 
\setcounter{Ceqindent}{0}\protect\begin{eqnarray}\hspace{-1.3ex}%
&\displaystyle\pi^\mu(\tau)\equiv m\hspace{-0.5mm}\protect\mbox{%
}\hspace{-0.1mm}\protect\int\protect\mbox{}\hspace{-0.5mm}\hspace
{-0.6mm}d^{\hspace{0.15ex}3}\hspace{-0.45mm}x\hspace{-0.5mm}%
\protect\mbox{}\hspace{-0.1mm}\protect\int_{-\infty}^{\infty}%
\protect\mbox{}\hspace{-0.5mm}\hspace{-0.6mm}d\tau\,\delta^{(4)}%
[x-z(\tau)]\,u^\mu(\tau)u^0(\tau),\protect\nonumber\setlength{%
\Cscr}{\value{CEht}\Ctenthex}\addtolength{\Cscr}{-1.0ex}\protect
\raisebox{0ex}[\value{CEht}\Ctenthex][\Cscr]{}\protect\end{eqnar%
ray}\setcounter{CEht}{10}where we write $\pi^\mu(\tau)$ on the u%
nderstanding that the given $\tau$ is the proper time of the par%
ticle at the corresponding co\-ordinate time $t$. The integratio%
n over all three-space can be performed immediately, with the th%
ree-part of the delta function simply yielding unity: \setcounter
{Ceqindent}{0}\protect\begin{eqnarray}\hspace{-1.3ex}&%
\displaystyle\pi^\mu(\tau)=m\hspace{-0.5mm}\protect\mbox{}\hspace
{-0.1mm}\protect\int_{-\infty}^{\infty}\protect\mbox{}\hspace{-0%
.5mm}\hspace{-0.6mm}d\tau\,\delta[t-z^0(\tau)]\,u^\mu(\tau)u^0(%
\tau).\protect\nonumber\setlength{\Cscr}{\value{CEht}\Ctenthex}%
\addtolength{\Cscr}{-1.0ex}\protect\raisebox{0ex}[\value{CEht}%
\Ctenthex][\Cscr]{}\protect\end{eqnarray}\setcounter{CEht}{10}We 
can now perform the integration over $\tau$ by noting the standa%
rd identity for a delta function of a function: \setcounter{Ceqi%
ndent}{0}\protect\begin{eqnarray}\hspace{-1.3ex}&\displaystyle
\delta[f(\tau)]=\sum_{\tau_z}\mbox{$\protect\displaystyle\protect
\frac{\delta(\tau-\tau_z)}{|df/d\tau(\tau_z)|}$},\protect
\nonumber\setlength{\Cscr}{\value{CEht}\Ctenthex}\addtolength{%
\Cscr}{-1.0ex}\protect\raisebox{0ex}[\value{CEht}\Ctenthex][\Cscr
]{}\protect\end{eqnarray}\setcounter{CEht}{10}where $\tau_z$ are 
the zeroes of $f(\tau)$. In this case we have \setcounter{Ceqind%
ent}{0}\protect\begin{eqnarray}f(\tau)=t-z^0(\tau),\protect
\nonumber\setlength{\Cscr}{\value{CEht}\Ctenthex}\addtolength{%
\Cscr}{-1.0ex}\protect\raisebox{0ex}[\value{CEht}\Ctenthex][\Cscr
]{}\protect\end{eqnarray}\setcounter{CEht}{10}whence \setcounter
{Ceqindent}{0}\protect\begin{eqnarray}\protect\left|\mbox{$%
\protect\displaystyle\protect\frac{df}{d\tau}$}\protect\right|=|%
u^0|,\protect\nonumber\setlength{\Cscr}{\value{CEht}\Ctenthex}%
\addtolength{\Cscr}{-1.0ex}\protect\raisebox{0ex}[\value{CEht}%
\Ctenthex][\Cscr]{}\protect\end{eqnarray}\setcounter{CEht}{10}an%
d hence \setcounter{Ceqindent}{0}\protect\begin{eqnarray}\hspace
{-1.3ex}&\displaystyle\pi^\mu=mu^\mu\mbox{$\protect\displaystyle
\protect\frac{u^0}{|u^0|}$},\protect\nonumber\setlength{\Cscr}{%
\value{CEht}\Ctenthex}\addtolength{\Cscr}{-1.0ex}\protect
\raisebox{0ex}[\value{CEht}\Ctenthex][\Cscr]{}\protect\end{eqnar%
ray}\setcounter{CEht}{10}where all quantities are assumed evalua%
ted at the given value of $\tau$ (or $t$). We now recognize the 
last factor as the \mbox{}\protect\/{\protect\em particle--antip%
article number\protect\/} $\beta$: \setcounter{Ceqindent}{0}%
\protect\begin{eqnarray}\hspace{-1.3ex}&\displaystyle\beta\equiv
\mbox{$\protect\displaystyle\protect\frac{u^0}{|u^0|}$}=\pm1,%
\protect\nonumber\setlength{\Cscr}{\value{CEht}\Ctenthex}%
\addtolength{\Cscr}{-1.0ex}\protect\raisebox{0ex}[\value{CEht}%
\Ctenthex][\Cscr]{}\protect\end{eqnarray}\setcounter{CEht}{10}an%
d hence \setcounter{Ceqindent}{0}\protect\begin{eqnarray}\protect
\left.\protect\begin{array}{rcl}\protect\displaystyle\pi^\mu=%
\beta mu^\mu=(m\gamma,m\gamma{\protect\mbox{\protect\boldmath{$v%
$}}}).\setlength{\Cscr}{\value{CEht}\Ctenthex}\addtolength{\Cscr
}{-1.0ex}\protect\raisebox{0ex}[\value{CEht}\Ctenthex][\Cscr]{}%
\protect\end{array}\protect\right.\protect\label{eq:NegEn-PiFina%
l}\protect\end{eqnarray}\setcounter{CEht}{10}We see that the ext%
ra factor of $\beta$ ``cancels'' the oddness of $u^\mu$ under ${%
\cal C}$, so that the four-vector $\pi^\mu$ is---like $T^{\mu\nu
}$ itself---\mbox{}\protect\/{\protect\em even\protect\/} under 
the ${\cal C}$ operation. \mbox{}\protect\/{\protect\em It is th%
e mechanical momenta\protect\/} $\pi^\mu$\mbox{}\protect\/{%
\protect\em, not the canonical momenta\protect\/} $p^\mu$\mbox{}%
\protect\/{\protect\em, that should be added together to compute 
the mechanical four-momentum of a system of particles.\protect\/%
} (Similarly, one must be careful to never confuse the \mbox{}%
\protect\/{\protect\em mechanical\protect\/} stress--energy tens%
or density $t^{\mu\nu}(x)$ with the \mbox{}\protect\/{\protect\em
canonical\protect\/} stress--energy tensor density ${\protect\it
\Theta\!\:}^{\mu\nu}(x)$ of field theory:$^{\ref{cit:Itzykson198%
0}}$ the former can be derived by functional differentiation of 
the Lagrangian with respect to the \mbox{}\protect\/{\protect\em
metric tensor\protect\/} $g_{\mu\nu}$, and is therefore always s%
ymmetric;$^{}$$^{\ref{cit:Misner1970}}$$^{,}$$^{\ref{cit:Belinfa%
nte1940}}${} the latter is defined in field theory with respect 
to the canonical momentum density, and in general possesses no p%
articular symmetry.)\par We can perform a similar integration of 
the \mbox{}\protect\/{\protect\em electromagnetic\protect\/} cur%
rent density vector $j^\mu(x)$, to obtain the ``effective flux o%
f charge'' $Q$ passing through a given spacelike hypersurface: 
\setcounter{Ceqindent}{0}\protect\begin{eqnarray}\hspace{-1.3ex}%
&\displaystyle Q(t)\equiv\hspace{-0.5mm}\protect\mbox{}\hspace{-%
0.1mm}\protect\int\protect\mbox{}\hspace{-0.5mm}\hspace{-0.6mm}d%
^{\hspace{0.15ex}3}\hspace{-0.45mm}\sigma_\mu\,j^\mu(t,{\protect
\mbox{\protect\boldmath{$x$}}}).\protect\nonumber\setlength{\Cscr
}{\value{CEht}\Ctenthex}\addtolength{\Cscr}{-1.0ex}\protect
\raisebox{0ex}[\value{CEht}\Ctenthex][\Cscr]{}\protect\end{eqnar%
ray}\setcounter{CEht}{10}By a similar analysis to that above, we 
simply find \setcounter{Ceqindent}{0}\protect\begin{eqnarray}%
\hspace{-1.3ex}&\displaystyle Q=\beta q,\protect\nonumber
\setlength{\Cscr}{\value{CEht}\Ctenthex}\addtolength{\Cscr}{-1.0%
ex}\protect\raisebox{0ex}[\value{CEht}\Ctenthex][\Cscr]{}\protect
\end{eqnarray}\setcounter{CEht}{10}so that $Q$ can be understood 
as the ``effective'' charge of the particle, if it had been in 
``normal'' particle motion. (The choice of what is ``particle'' 
motion and what is ``antiparticle'' motion is of course implicit%
ly contained in the \mbox{}\protect\/{\protect\em direction%
\protect\/} of the timelike four-normal $n^\mu$.)\par
\refstepcounter{section}\vspace{1.5\baselineskip}\par{\centering
\bf\thesection. Conclusions\\*[0.5\baselineskip]}\protect\indent
\label{sect:Conclusions}\begin{table}[tbp]\small\begin{centre}%
\begin{tabular}{lccccl} Quantity & Symbol & Rank & ${\cal C}$ & 
${\cal P}{\cal T}$ & Value \\ \hline Mass & $m$ & 0 & $+$ & $+$ 
& Free parameter \\ Electric charge & $q$ & 0 & $+$ & $+$ & Free 
parameter \\ Proper time & $\tau$ & 0 & $-$ & $+$ & $d\tau=\beta
\,dt/\gamma(t)$ \\ Theta time & $\theta$ & 0 & $-$ & $+$ & $d%
\theta=d\tau/m(\tau)$ \\ Particle number & $\beta$ & 0 & $-$ & $%
+$ & $u^{0\!}/|u^0|=u^{0\!}/\gamma={\pm1}$ \\ Lagrangian & $L$ & 
0 & $+$ & $+$ & $T^{\mu\nu\!}g_{\mu\nu}/2+J^\mu\!A_\mu$ \\ Hamil%
tonian & ${\cal H}$ & $0$ & $+$ & $+$ & $p_\mu u^\mu-L=(p-qA)^{2%
\!}/2m$ \\ Effective charge & $Q$ & $0$ & $-$ & $+$ & $\beta q$ 
\\ \hline Position & $z^\mu$ & 1 & $+$ & $-$ & $z^\mu(\tau)$, st%
ate vector\\ Velocity & $u^\mu$ & 1 & $-$ & $-$ & $dz^{\mu\!}/d%
\tau=(\beta\gamma,\beta\gamma{\protect\mbox{\protect\boldmath{$v%
$}}})$ \\ Canonical momentum & $p^\mu$ & 1 & $-$ & $-$ & $%
\partial L/\partial u_\mu=mu^\mu+qA^\mu$ \\ Electromagnetic pote%
ntial & $A^\mu$ & 1 & $-$ & $-$ & $\partial_\mu\partial^\mu\!A_%
\nu-\partial_\mu\partial_\nu A^\mu=j_\nu$ \\ Electromagnetic cur%
rent & $J^\mu$ & 1 & $-$ & $-$ & $qu^\mu$ \\ Mechanical momentum 
& $\pi^\mu$ & 1 & $+$ & $-$ & $\beta mu^\mu=(m\gamma,m\gamma{%
\protect\mbox{\protect\boldmath{$v$}}})$ \\ \hline Electromagnet%
ic field & $F^{\mu\nu}$ & 2 & $-$ & $+$ & $\partial^\mu\!A^{\nu
\!}-\partial^\nu\!A^\mu$ \\ Mechanical stress--energy & $T^{\mu
\nu}$ & 2 & $+$ & $+$ & $mu^\mu u^\nu$ \\ \hline Mechanical angu%
lar momentum & $M^{\mu\nu\rho}$ & 3 & $+$ & $-$ & $z^\nu T^{\mu
\rho}-z^\rho T^{\mu\nu}$ \\ \end{tabular} \protect\mbox{}\vspace
{-5.5ex}\vspace{\CZtbldist}\\\end{centre}\protect\caption[Error]%
{\protect\label{tbl:Conclusions-Invariants}Lorentz-covariant qua%
ntities for a classical point charge.}\protect\end{table}{}The S%
tueckelberg--Feynman picture of antiparticles being simply parti%
cles ``moving backwards in proper time'' can be seen to be an in%
tegral and important part of relativistic classical mechanics, w%
hich only requires minor additions to standard texts on special 
relativity. Furthermore, the historical misconceptions of the ``%
negative energy problem'' in relativistic quantum mechanics can 
be avoided by a thorough understanding of the difference between 
the \mbox{}\protect\/{\protect\em canonical\protect\/} momentum 
$p^\mu$ of Lagrangian theory, and the ``mechanical'' momentum $%
\pi^\mu$ that dictates the kinematical and gravitational propert%
ies of an object.\par Finally, we summarize the properties of th%
e various important Lorentz-covariant quantities for a classical 
point charge in Table~\protect\ref{tbl:Conclusions-Invariants}.%
\par\vspace{1.5\baselineskip}\par{\centering\bf Acknowledgments%
\\*[0.5\baselineskip]}\protect\indent Helpful discussions with S%
.~A.~Wout\-huysen, J.~W.~G. Wig\-nall, A.~G.~Klein, G.~I.~Opat, 
and M.~J.~Thom\-son are gratefully acknowledged. This work was s%
upported in part by the Australian Research Council. We were sad%
dened to learn of the death of Prof.~Wout\-huysen\ during the co%
mpletion of this work.\par\vspace{1.5\baselineskip}\par{%
\centering\bf References\\*[0.5\baselineskip]}\protect\mbox{}%
\vspace{-\baselineskip}\vspace{-2ex}\settowidth\CGDnum{[\ref{cit%
last}]}\setlength{\CGDtext}{\textwidth}\addtolength{\CGDtext}{-%
\CGDnum}\begin{list}{Error!}{\setlength{\labelwidth}{\CGDnum}%
\setlength{\labelsep}{0.75ex}\setlength{\leftmargin}{0ex}%
\setlength{\rightmargin}{0ex}\setlength{\itemsep}{0ex}\setlength
{\parsep}{0ex}}\protect\frenchspacing\setcounter{CBtnc}{1}\item[%
{\hfill\makebox[0ex][r]{\raisebox{0ex}[1ex][0ex]{$^{\mbox{$%
\fnsymbol{CBtnc}$}}$}}}]\addtocounter{CBtnc}{1}jpc@physics.unime%
lb.edu.au; http:/$\!$/www.ph.unimelb.edu.au/$\sim$jpc.\item[{%
\hfill\makebox[0ex][r]{\raisebox{0ex}[1ex][0ex]{$^{\mbox{$%
\fnsymbol{CBtnc}$}}$}}}]\addtocounter{CBtnc}{1}mckellar@physics.%
unimelb.edu.au.\item[{\hfill\makebox[0ex][r]{\raisebox{0ex}[1ex]%
[0ex]{$^{\mbox{$\fnsymbol{CBtnc}$}}$}}}]\addtocounter{CBtnc}{1}a%
rawlins@physics.unimelb.edu.au.\addtocounter{CBcit}{1}\item[%
\hfill$^{\arabic{CBcit}}$]\renewcommand\theCscr{\arabic{CBcit}}%
\protect\refstepcounter{Cscr}\protect\label{cit:Costella1995}J.~%
P.~Costella\ and B.~H.~J.~McKellar, \renewcommand\theCscr{Costel%
la\ and McKellar}\protect\refstepcounter{Cscr}\protect\label{au:%
Costella1995}\renewcommand\theCscr{1995}\protect\refstepcounter{%
Cscr}\protect\label{yr:Costella1995}``The Foldy--Wouthuysen tran%
sformation,'' Am. J. Phys.\ {\bf63}, 1119--1121\ (1995).%
\addtocounter{CBcit}{1}\item[\hfill$^{\arabic{CBcit}}$]%
\renewcommand\theCscr{\arabic{CBcit}}\protect\refstepcounter{Csc%
r}\protect\label{cit:Newton1949}T.~D.~Newton\ and E.~P.~Wigner, 
\renewcommand\theCscr{Newton\ and Wigner}\protect\refstepcounter
{Cscr}\protect\label{au:Newton1949}\renewcommand\theCscr{1949}%
\protect\refstepcounter{Cscr}\protect\label{yr:Newton1949}``Loca%
lized states for elementary systems,'' Rev. Mod. Phys.\ {\bf21}, 
400--406\ (1949).\addtocounter{CBcit}{1}\item[\hfill$^{\arabic{C%
Bcit}}$]\renewcommand\theCscr{\arabic{CBcit}}\protect
\refstepcounter{Cscr}\protect\label{cit:Foldy1950}L.~L.~Foldy\ a%
nd S.~A.~Wouthuysen, \renewcommand\theCscr{Foldy\ and Wouthuysen%
}\protect\refstepcounter{Cscr}\protect\label{au:Foldy1950}%
\renewcommand\theCscr{1950}\protect\refstepcounter{Cscr}\protect
\label{yr:Foldy1950}``On the Dirac theory of spin $1/2$ particle%
s and its non-relativistic limit,'' Phys. Rev.\ {\bf78}, 29--36\ 
(1950).\addtocounter{CBcit}{1}\item[\hfill$^{\arabic{CBcit}}$]%
\renewcommand\theCscr{\arabic{CBcit}}\protect\refstepcounter{Csc%
r}\protect\label{cit:Jackson1975}J.~D.~Jackson, \renewcommand
\theCscr{Jackson}\protect\refstepcounter{Cscr}\protect\label{au:%
Jackson1975}\renewcommand\theCscr{1975}\protect\refstepcounter{C%
scr}\protect\label{yr:Jackson1975}\mbox{}\protect\/{\protect\em
Classical Electrodynamics\protect\/}, 2nd~ed. (Wiley, New York, 
1975).\addtocounter{CBcit}{1}\item[\hfill$^{\arabic{CBcit}}$]%
\renewcommand\theCscr{\arabic{CBcit}}\protect\refstepcounter{Csc%
r}\protect\label{cit:Stueckelberg1942}E.~C.~G.~Stueckelberg, 
\renewcommand\theCscr{Stueckelberg}\protect\refstepcounter{Cscr}%
\protect\label{au:Stueckelberg1942}\renewcommand\theCscr{1942}%
\protect\refstepcounter{Cscr}\protect\label{yr:Stueckelberg1942}%
``La m\'ecanique du point mat\'eriel en th\'eorie del relativit%
\'e et en th\'eorie des quanta,'' Helv. Phys. Act.\ {\bf15}, 23%
--37\ (1942).\addtocounter{CBcit}{1}\item[\hfill$^{\arabic{CBcit%
}}$]\renewcommand\theCscr{\arabic{CBcit}}\protect\refstepcounter
{Cscr}\protect\label{cit:Feynman1948}R.~P.~Feynman, \renewcommand
\theCscr{Feynman}\protect\refstepcounter{Cscr}\protect\label{au:%
 Feynman1948}\renewcommand\theCscr{1948}\protect\refstepcounter{%
Cscr}\protect\label{yr:Feynman1948}``A relativisitic cut-off for 
classical electrodynamics,'' Phys. Rev.\ {\bf74}, 939--946\ (194%
8).\addtocounter{CBcit}{1}\item[\hfill$^{\arabic{CBcit}}$]%
\renewcommand\theCscr{\arabic{CBcit}}\protect\refstepcounter{Csc%
r}\protect\label{cit:Feynman1949}R.~P.~Feynman, \renewcommand
\theCscr{Feynman}\protect\refstepcounter{Cscr}\protect\label{au:%
 Feynman1949}\renewcommand\theCscr{1949}\protect\refstepcounter{%
Cscr}\protect\label{yr:Feynman1949}``The theory of positrons,'' 
Phys. Rev.\ {\bf76}, 749--759\ (1949).\addtocounter{CBcit}{1}%
\item[\hfill$^{\arabic{CBcit}}$]\renewcommand\theCscr{\arabic{CB%
cit}}\protect\refstepcounter{Cscr}\protect\label{cit:Feynman1987%
}R.~P.~Feynman, \renewcommand\theCscr{Feynman}\protect
\refstepcounter{Cscr}\protect\label{au:Feynman1987}\renewcommand
\theCscr{1987}\protect\refstepcounter{Cscr}\protect\label{yr:Fey%
nman1987}``The reason for antiparticles,'' in \mbox{}\protect\/{%
\protect\em Elementary Particles and the Laws of Physics: The 19%
86 Dirac Memorial Lectures\protect\/} (Cambridge Univ. Press, Ca%
mbridge, 1987).\addtocounter{CBcit}{1}\item[\hfill$^{\arabic{CBc%
it}}$]\renewcommand\theCscr{\arabic{CBcit}}\protect
\refstepcounter{Cscr}\protect\label{cit:Misner1970}C.~W.~Misner, 
K.~S.~Thorne, and J.~A.~Wheeler, \renewcommand\theCscr{Misner, T%
horne, and Wheeler}\protect\refstepcounter{Cscr}\protect\label{a%
u:Misner1970}\renewcommand\theCscr{1970}\protect\refstepcounter{%
Cscr}\protect\label{yr:Misner1970}\mbox{}\protect\/{\protect\em
Gravitation\protect\/} (Freeman, New York, 1970).\addtocounter{C%
Bcit}{1}\item[\hfill$^{\arabic{CBcit}}$]\renewcommand\theCscr{%
\arabic{CBcit}}\protect\refstepcounter{Cscr}\protect\label{cit:G%
oldstein1980}H.~Goldstein, \renewcommand\theCscr{Goldstein}%
\protect\refstepcounter{Cscr}\protect\label{au:Goldstein1980}%
\renewcommand\theCscr{1980}\protect\refstepcounter{Cscr}\protect
\label{yr:Goldstein1980}\mbox{}\protect\/{\protect\em Classical 
Mechanics\protect\/}, 2nd~ed. (Addison-Wesley, Massachusetts, 19%
80).\addtocounter{CBcit}{1}\item[\hfill$^{\arabic{CBcit}}$]%
\renewcommand\theCscr{\arabic{CBcit}}\protect\refstepcounter{Csc%
r}\protect\label{cit:Costella1994}J.~P.~Costella, \renewcommand
\theCscr{Costella}\protect\refstepcounter{Cscr}\protect\label{au%
:Costella1994}\renewcommand\theCscr{1994}\protect\refstepcounter
{Cscr}\protect\label{yr:Costella1994}Ph.D.\ thesis, The Universi%
ty of Melbourne\ (1994), unpublished; available from the author'%
s home page, listed above.\addtocounter{CBcit}{1}\item[\hfill$^{%
\arabic{CBcit}}$]\renewcommand\theCscr{\arabic{CBcit}}\protect
\refstepcounter{Cscr}\protect\label{cit:Itzykson1980}C.~Itzykson%
\ and J.-B.~Zuber, \renewcommand\theCscr{Itzykson\ and Zuber}%
\protect\refstepcounter{Cscr}\protect\label{au:Itzykson1980}%
\renewcommand\theCscr{1980}\protect\refstepcounter{Cscr}\protect
\label{yr:Itzykson1980}\mbox{}\protect\/{\protect\em Quantum Fie%
ld Theory\protect\/} (McGraw-Hill, New York, 1980).\addtocounter
{CBcit}{1}\item[\hfill$^{\arabic{CBcit}}$]\renewcommand\theCscr{%
\arabic{CBcit}}\protect\refstepcounter{Cscr}\protect\label{cit:B%
elinfante1940}F.~Belinfante, \renewcommand\theCscr{Belinfante}%
\protect\refstepcounter{Cscr}\protect\label{au:Belinfante1940}%
\renewcommand\theCscr{1940}\protect\refstepcounter{Cscr}\protect
\label{yr:Belinfante1940}``On the current and the density of the 
electric charge, the energy, the linear momentum, and the angula%
r momentum of arbitrary fields,'' Physica\ {\bf7}, 449--474\ (19%
40).\renewcommand\theCscr{\arabic{CBcit}}\protect\refstepcounter
{Cscr}\protect\label{citlast}\settowidth\Cscr{$^{\ref{cit:Belinf%
ante1940}}$}\end{list}\par\end{document}